\documentclass[%
 reprint,
 amsmath,amssymb,
 aps,
 prl,
]{revtex4-2}

\usepackage{graphicx}
\usepackage{dcolumn}
\usepackage{bm}

\usepackage{graphicx}
\usepackage[normalem]{ulem}
\usepackage{color,bm}
\usepackage{amsmath,amssymb}
\usepackage[T1]{fontenc}
\usepackage{siunitx}
\usepackage{url}

\begin{document}



\title{Probing van der Waals interactions and detecting polar molecules by F\"orster resonance energy transfer with Rydberg atoms at temperatures below 100~mK}

\author{J. Zou }
\author{S. D. Hogan}%
\affiliation{Department of Physics and Astronomy, University College London, Gower Street, London WC1E 6BT, UK}%

\date{\today}

\begin{abstract}

Electric-field-controlled F\"orster resonance energy transfer (FRET) between Rydberg helium (He) atoms and ground-state ammonia (NH$_3$) molecules has been studied at translational temperatures below 100~mK. The experiments were performed in an intrabeam collision apparatus with pulsed supersonic beams of NH$_3$ seeded in He. A range of F\"orster resonances, between triplet Rydberg states in He with principal quantum numbers of 38, 39 and 40, and the inversion intervals in NH$_3$ were investigated. Resonance widths as low as $100\pm20$~MHz were observed for Rydberg-Rydberg transitions with electric dipole transition moments of 3270~D. These widths result from binary collisions at a mean center-of-mass speed of $19.3\pm2.6$~m/s. For transitions in which the initially prepared Rydberg states were strongly polarized, with large induced static electric dipole moments, van der Waals interactions between the collision partners increased the resonance widths to $\sim750$~MHz. From measurements of the rate of FRET for the narrowest resonances observed, a density of NH$_3$ of $(9.4\pm0.3)\times10^{9}$~cm$^{-3}$ in the upper ground-state inversion sublevel in the interaction region of the apparatus was determined non-destructively. 
\end{abstract}

\maketitle


\section{Introduction}

Investigations of gas-phase chemical dynamics with neutral atoms and molecules at low collision energies, $E_{\rm{kin}}$, or temperatures, i.e., $E_{\rm{kin}} / k_{\rm{B}}\lesssim1$~K, offer opportunities to study effects of de Broglie wave interference~\cite{zhou21a}, scattering resonances~\cite{paliwal21a}, correlated excitations~\cite{gao18a}, stereo-dynamics~\cite{zou19a} and long-range interactions~\cite{gawlas20a} in exquisite detail and with extreme quantum state selectivity. At these low temperatures, van der Waals interactions can play an important role at long range. These include the $1/R^3$ ‘dipole-dipole' interaction, and the $1/R^6$ ‘dispersion' interaction~\cite{muller94a}.  For systems with allowed electric dipole transitions at similar frequencies, resonant dipole-dipole interactions can also occur and give rise to F\"orster resonance energy transfer (FRET)~\cite{perrin32a,forster46a,jones19a}. 

The dipole-dipole interaction energy, $V_{\mathrm{dd}}$, between particles A and B, can be expressed as~\cite{gallagher08a}
\begin{equation}\label{eq:Vdd}
V_{\rm{dd}}(\vec{R}) = \frac{1}{4 \pi \epsilon_0} \left[\frac{\mu_{\rm{A}}\mu_{\rm{B}}}{R^3} - 3 \frac{(\vec{\mu}_{\rm{A}} \cdot \vec{R})(\vec{\mu}_{\rm{B}} \cdot \vec{R})}{R^5}\right],
\end{equation}
where $\mu_{\mathrm{A}} = |\vec{\mu}_{\mathrm{A}}|$ ($\mu_{\mathrm{B}} = |\vec{\mu}_{\mathrm{B}}|$) is the electric dipole moment of particle A (B), $R = |\vec{R}|$ is the interparticle distance, and $\epsilon_0$ is the vacuum permittivity. For the $1/R^3$ van der Waals interaction, the dipole moment $\vec{\mu}_{\mathrm{A}}$ ($\vec{\mu}_{\mathrm{B}}$) is the induced static electric dipole moment of particle A (B), i.e., $\vec{\mu}_{\mathrm{A}\,(\mathrm{B})}^{\mathrm{\,static}} = -\mathrm{d}W_{\mathrm{Stark,\,A\,(\mathrm{B})}}/\mathrm{d}F$, the derivative of the Stark shift, where $W_{\mathrm{Stark},\,\mathrm{A}\,(\mathrm{B})}$ is the Stark energy and $F$ is the electric field strength~\cite{zhelyazkova16a}. For the resonant dipole-dipole interaction associated with FRET, the dipole moments $\vec{\mu}_{\mathrm{A\,(B)}}^{\mathrm{\,trans}} = \langle \mathrm{f}_{\mathrm{A\,(\mathrm{B})}} |e\hat{r}| \mathrm{i}_{\mathrm{A\,(B)}}\rangle$, are the electric dipole transition moments between initial and final states, $|\mathrm{i}_{\mathrm{A\,(B)}}\rangle$ and $|\mathrm{f}_{\mathrm{A\,(B)}}\rangle$ in particle A (B), respectively. 

These dipole-dipole interactions play important roles, e.g., in intra-molecular Coulombic decay~\cite{cederbaum97a,janke15a}, energy transfer in light harvesting complexes~\cite{mirkovic17a}, and quantum simulation~\cite{blackmore19a}. If one of the interacting particles is in a Rydberg state with a high principal quantum number $n$, these interactions can be particularly pronounced. This is because the static electric dipole moments of, and electric dipole transition moments between pairs of Rydberg states scale with $n^2$, and for $n\gtrsim30$ are $\gtrsim1000$~D~\cite{gallagher94a}. Dipole-dipole interactions between polar ground-state molecules, and atoms or molecules in Rydberg states are central to the investigation of dipole-bound states~\cite{farooqi03a}, and studies of how long-range interactions can be exploited to regulate access to shorter range chemistry~\cite{allmendinger16a}. They also have applications in coherent control~\cite{kuznetsova11a}, sympathetic cooling~\cite{huber12a,zhao12a}, and non-destructive detection~\cite{zeppenfeld17a}. 

Early investigations of FRET in collisions of Rydberg atoms with polar ground-state molecules were performed with room temperature gases~\cite{smith78a,stebbings82a,petitjean84a,petitjean86a,ling93a,zhelyazkova17a,zhelyazkova17b,jarisch18a}. Recently studies of the effects of electric fields on FRET between Rydberg He atoms and ground-state NH$_3$ were extended to collision energies $E_{\rm{kin}}/k_{\mathrm{B}}<1$~K using an intrabeam collision apparatus~\cite{gawlas20a}. The work reported here builds on these pioneering experiments by using FRET as a tool to probe van der Waals interactions between the atoms and the molecules, and to non-destructively determine molecule number densities at translational temperatures $<100$~mK. 


\begin{figure}
\includegraphics[width=0.9\columnwidth]{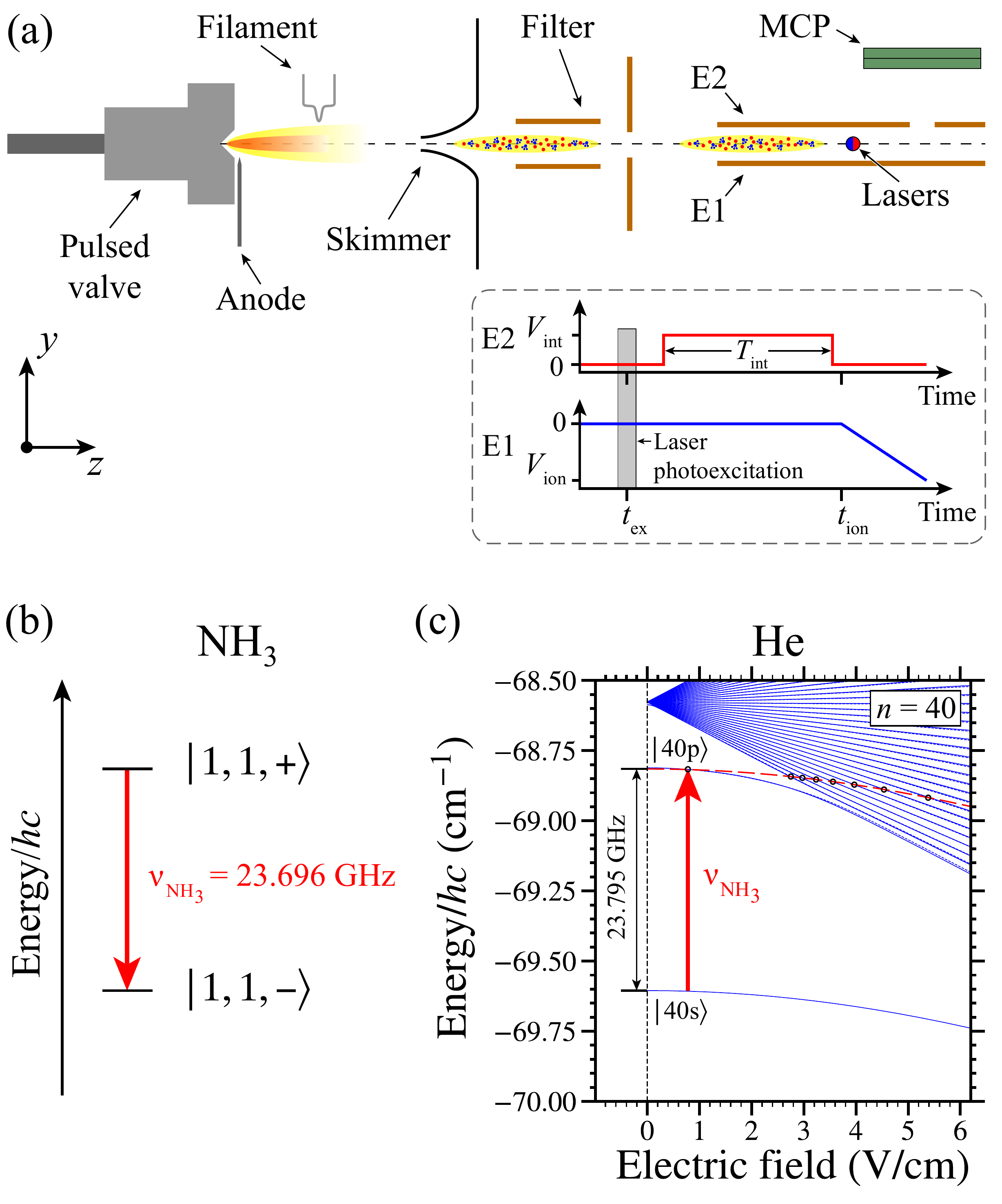}
\caption{(a) Schematic diagram of the experimental apparatus. Energy-level diagram of (b) the ground-state inversion interval in NH$_3$, and (c) triplet Rydberg states in He. The Stark map in (c) contains sub-levels for which $m_{\ell} = 0$ and~1 (continuous and dashed blue curves, respectively). The red dashed curve represents the energy of the $|40\rm{s}\rangle$ state offset by $\nu_{\mathrm{NH}_3}$.}\label{fig1}
\end{figure}

\section{Experiment}

The experiments were performed in an intrabeam collision apparatus~\cite{amarasinghe17a,gawlas20a}. This is depicted schematically in Figure~\ref{fig1}(a). A supersonic beam of NH$_3$ seeded in He (1:99 by pressure), was generated using a pulsed valve. An electric discharge at the exit of the valve populated the metastable 1s2s $^3$S$_1$ level in He~\cite{halfmann00a}. After passing through a skimmer and charged particle filter, the beam entered between two parallel electrodes, E1 and E2. There, Rydberg-state photoexcitation occurred at time $t_{\mathrm{ex}}$ [inset in Figure~\ref{fig1}(a)] using co-propagating CW laser beams at wavelengths of 388.975~nm, and between 789.333 and 788.861~nm to drive each step of the 1s2s\,$\rm{^3S_1} \rightarrow$ 1s3p\,$\rm{^3P_2} \rightarrow$ 1s$n$s\,$\rm{^3S_1}$ two-photon excitation scheme, respectively~\cite{hogan18a}. Pulsed potentials with amplitude $V_{\mathrm{int}}$ and duration $T_{\rm{int}}$ were applied to E2 after photoexcitation to generate electric fields of up to 8~V/cm and induce Stark shifts of transitions between the Rydberg states. At time $t_{\mathrm{ion}}\simeq12~\mu$s, quantum-state-selective detection of the Rydberg atoms was performed by pulsed electric field ionization. Ionized electrons were accelerated through an aperture in E2 and collected at a microchannel plate (MCP) detector.

\begin{figure}
\includegraphics[width=0.99\columnwidth]{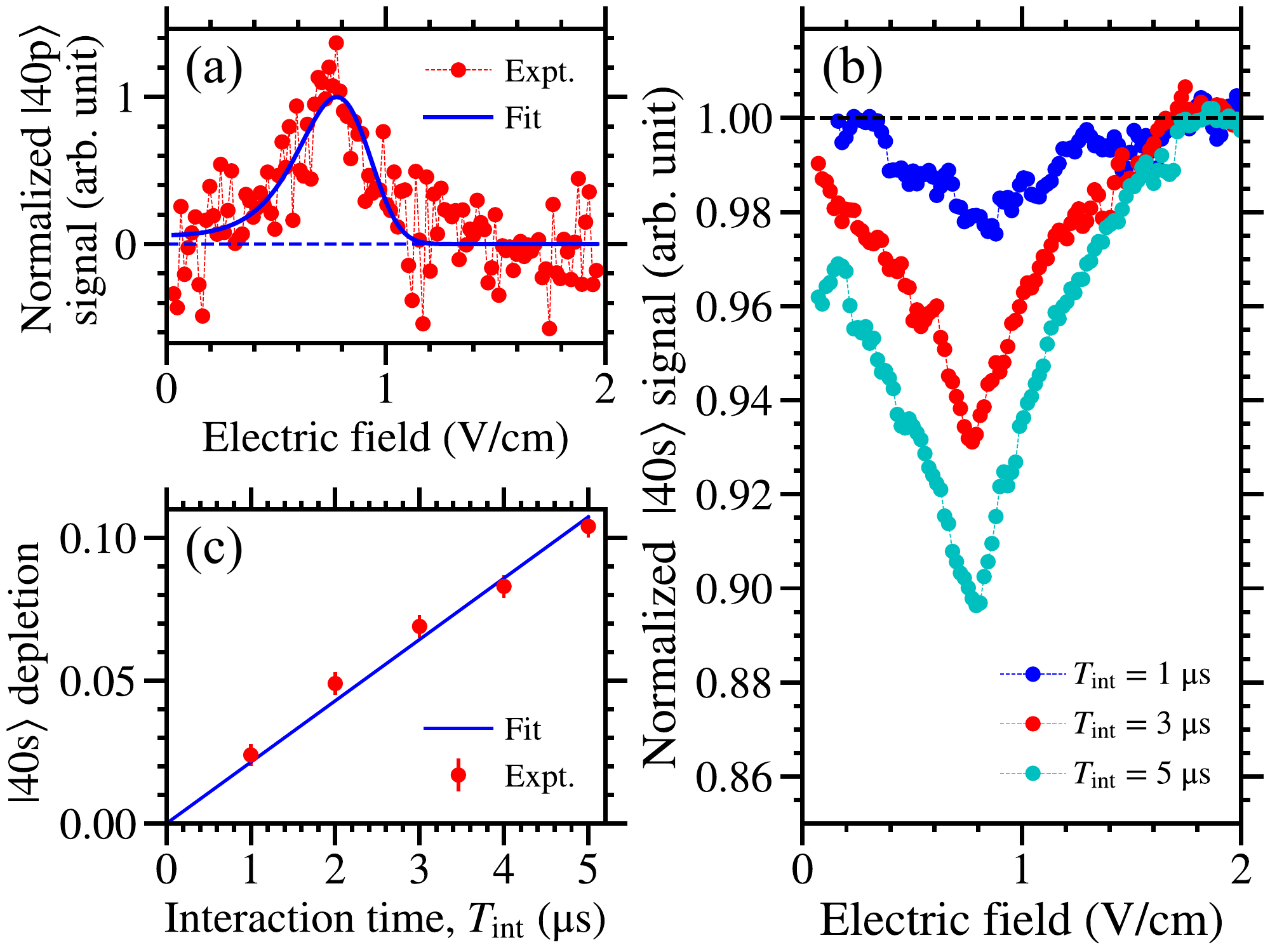}
\caption{(a) $|40\rm{p}\rangle$-electron signal recorded after initial population of the $|40\rm{s}\rangle$ state and $T_{\rm{int}}=2~\mu$s (red points). The continuous blue curve represents a Gaussian function with a FWHM of $100\pm20$~MHz fit to the experimental data in the frequency domain. (b) Depletion of the $|40\rm{s}\rangle$-electron signal for $T_{\rm{int}}=1$, 3 and $5~\mu$s. (c) Dependence of the depletion of the $|40\rm{s}\rangle$-electron signal on resonance (red points) on the value of $T_{\rm{int}}$. The slope of the line fit to the full set of data in (c) yields a depletion rate of $\Gamma_{\mathrm{dep},\,|40\mathrm{s}\rangle} = 21.5\pm0.7$~kHz.}\label{fig2}
\end{figure}

NH$_3$ in the X$\,^1$A$_1$ ground electronic state, and He in triplet Rydberg states are well suited to studies of FRET from nuclear to electronic degrees of freedom. In the $J=K=1$ rotational state ($J$ and $K$ are the rotational quantum number, and the projection of the rotational angular momentum vector onto the symmetry axis of the molecule, respectively), with vibrational quantum number $v=0$, the symmetric, $|J = 1,K = 1,+\rangle$, and antisymmetric, $|1,1,-\rangle$, inversion sublevels in NH$_3$ are separated by $\nu_{\mathrm{NH}_3} = 23.696$~GHz [Figure~\ref{fig1}(b)]. The electric dipole moment of the molecule in these states is $\mu_{\mathrm{NH}_3} = 1.47$~D. In zero field, the interval between the 1s40s$\,^3$S$_1$ ($|40\mathrm{s}\rangle$; quantum defect $\delta_{40\mathrm{s}} =  0.296681$) and 1s40p$\,^3$P$_J$ ($|40\mathrm{p}\rangle$; $\delta_{40\mathrm{p}} =  0.068349$) levels in He is $\nu_{40\mathrm{s},40\mathrm{p}} = 23.794$~GHz [Figure~\ref{fig2}(c)]~\cite{drake99a}, and the $|40\mathrm{s}\rangle\rightarrow|40\mathrm{p}\rangle$ $\Delta m_{\ell} = 1$ electric dipole transition moment is $\mu_{40\mathrm{p},40\mathrm{s}}^{\mathrm{trans}} = 3300$~D ($m_{\ell}$ is the azimuthal quantum number). These Rydberg states exhibit quadratic Stark shifts in weak electric fields. The static electric dipole polarizability of the $|40\mathrm{s}\rangle$ [$|40\mathrm{p}\rangle$] state with $m_{\ell} = 0$ [$m_{\ell} = 1$] is 217~MHz/(V/cm)$^2$ [531~MHz/(V/cm)$^2$]. Consequently, the $|40\mathrm{s}\rangle\rightarrow|40\mathrm{p}\rangle$ transition shifts to a lower wave number as the electric field strength increases, and is resonant with $\nu_{\mathrm{NH}_3}$ in a field of 0.78~V/cm. This is seen by comparing the energy of the $|40\mathrm{p}\rangle$ state with $m_{\ell} = 1$ in Figure~\ref{fig1}(c) with the dashed red curve offset from the $|40\mathrm{s}\rangle$ state by $\nu_{\mathrm{NH}_3}$. At this F\"orster resonance, the electric dipole moment associated with the $\Delta m_{\ell}=1$ transition between the $\ell$-mixed $|40\mathrm{s}'\rangle$ and $|40\mathrm{p}'\rangle$ states is $\mu_{40\mathrm{p}',40\mathrm{s}'}^{\mathrm{trans}} = 3270$~D. The 1s40p\,$^3$P$_J$ fine-structure in He is $\sim3.2$~MHz~\cite{deller18a}, and the Stark shifts of the $|1,1,\pm\rangle$ states in NH$_3$ are $<500$~Hz in electric fields up to 10~V/cm -- these effects are neglected in the analysis of the current experiments. 

\section{Results}

\subsection{Collision speed and molecule number density}

The electric-field-dependence of the FRET process for He atoms prepared in the $|40\mathrm{s}\rangle$ state that underwent collisions with NH$_3$ in the $|1,1,+\rangle$ state is shown in Figure~\ref{fig2}. When recording the data in Figure~\ref{fig2}(a), the $|40\mathrm{p}\rangle$ signal was monitored as $V_{\mathrm{int}}$ was adjusted for $T_{\mathrm{int}} = 2~\mu$s. In this spectrum, resonant population transfer to the $|40\mathrm{p}\rangle$ state occurs in a field close to 0.78~V/cm. The full-width-at-half-maximum (FWHM) of the resonance was determined by transforming from the electric-field to the frequency domain using the calculated Stark shifts of the Rydberg states, and fitting a Gaussian function (continuous blue curve). In this process only the amplitude and FWHM of the Gaussian were free fit parameters. The resonance frequency was set to $\nu_{\mathrm{NH}_3}$. The resulting FWHM resonance width was $100\pm20$~MHz. 

FRET widths for binary interactions in the gas phase, can be interpreted by considering a mean center-of-mass collision speed $\overline{v}$, and an impact parameter $b$~\cite{gallagher92a}. For the average resonant dipole-dipole interaction energy between a He Rydberg atom and a randomly oriented NH$_3$ molecule, the collision-time-limited resonance width is~\cite{gallagher92a,zhelyazkova17b}
\begin{equation}
\Gamma_{\mathrm{int}} = \frac{\Delta E_{\rm{FWHM}}}{h} \backsimeq \frac{\overline{v}}{b} = \sqrt{\frac{4 \pi \epsilon_0 \overline{v}^3 h}{\mu_{\rm{He}}\,\mu_{\rm{NH_3}}}},\label{eq:width}
\end{equation}
where $h$ is Planck's constant. Since the data in Figure~\ref{fig2}(a) were recorded for the isolated $|40\mathrm{s}'\rangle\rightarrow|40\mathrm{p}'\rangle$ transition, the value of $\overline{v}$ could be determined from the measured FWHM using this expression to be $19.3\pm2.6$~m/s. This corresponds to a mean center-of-mass collision energy of $E_{\mathrm{kin}}/k_{\mathrm{B}} = 73\pm20$~mK. The assumption in this analysis, that the width of the resonance in Figure~\ref{fig2}(a) is solely a result of the finite atom-molecule collision time, means that this collision speed and collision energy represent upper bounds on those in the experiments. From previous intrabeam collision studies in pulsed supersonic beams~\cite{amarasinghe17a}, we estimate this collision speed of 19.3~m/s is at most an overestimate of the true value by a factor of $\sim2$. 


The rate of population transfer by FRET in the experiments was determined by monitoring the depletion of the population of the $|40\mathrm{s}\rangle$ state on the $|40\mathrm{s}'\rangle\rightarrow|40\mathrm{p}'\rangle$ resonance, for a range of values of $T_{\mathrm{int}}$ [see Figure~\ref{fig2}(b)]. From these data, the fractional depletion [red points in Figure~\ref{fig2}(c)] was obtained, and depends linearly on $T_{\mathrm{int}}$. The slope of the line fit to these data, yields a depletion rate of $\Gamma_{\mathrm{dep},\,|40\mathrm{s}\rangle} = 21.5\pm0.7$~kHz. Considering only pseudo-first-order kinetics, the reaction rate $\Gamma_{\mathrm{FRET}} = \overline{v}\,\sigma_{\mathrm{FRET}}\,N_{\mathrm{NH}_3}$, where $\sigma_{\mathrm{FRET}}$ and $N_{\mathrm{NH}_3}$ are the FRET cross-section and the number density of NH$_3$ in the upper inversion sublevel, respectively. Since $\sigma_{\mathrm{FRET}} = \pi\,b^2 = \mu_{\rm{He}}\,\mu_{\rm{NH_3}}/(4\epsilon_0 v h)$~\cite{gallagher92a}, $N_{\mathrm{NH}_3}$ was determined from $\Gamma_{\mathrm{dep},\,|40\mathrm{s}\rangle}$ to be $(9.4\pm0.3)\times10^9$~cm$^{-3}$. The linearity of the data across the full range of interaction times in Fig.~\ref{fig2}(c) indicates that the assumption of pseudo-first-order kinetics is reasonable. However, we note that if only measurements up to $T_{\mathrm{int}} = 2$~$\mu$s were considered, the density determined would increase by 15\%. 

\begin{figure}
\includegraphics[width=0.6\columnwidth]{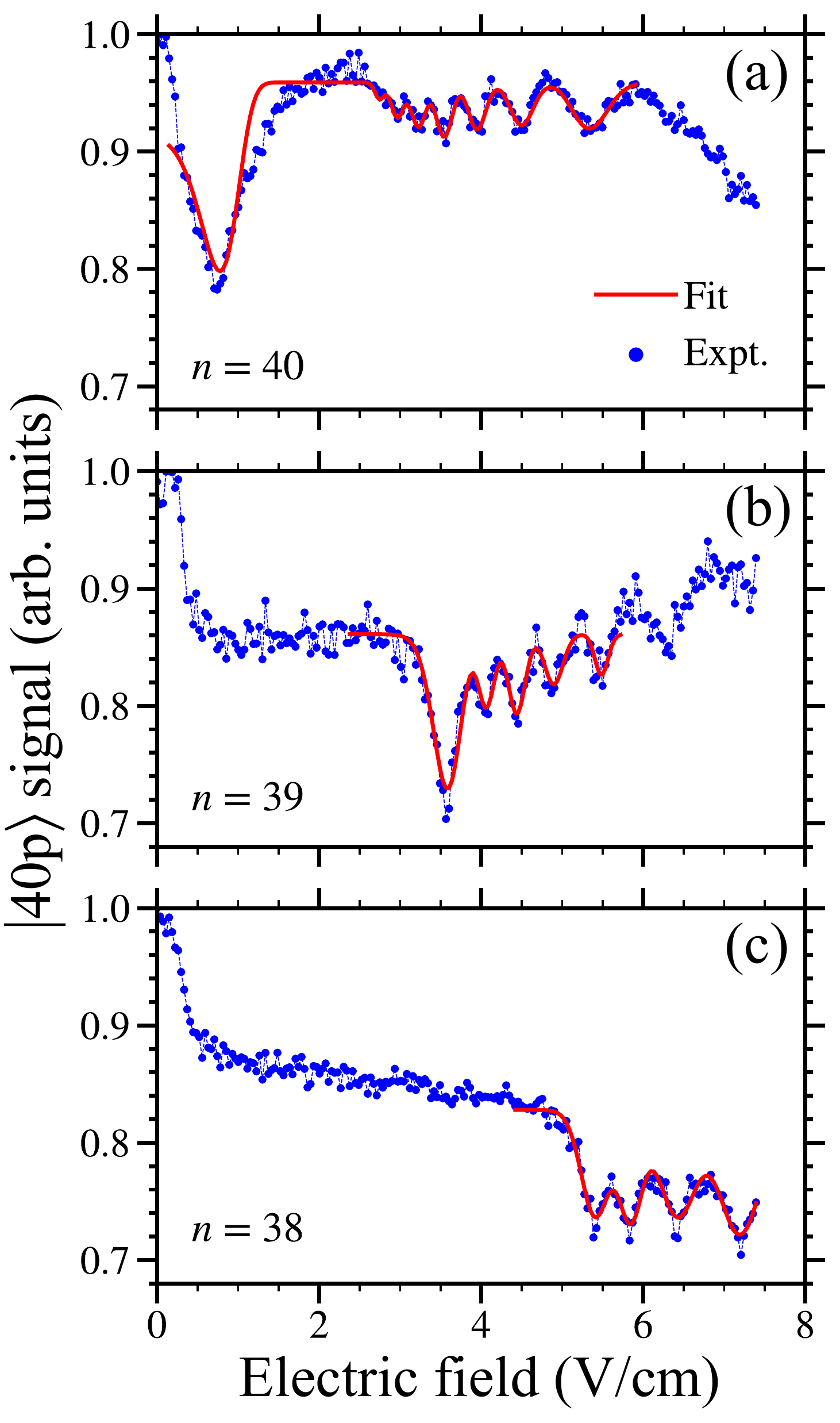}
\caption{FRET monitored by depletion of the population of $|n\mathrm{s}\rangle$ Rydberg states for (a) $n=38$, (b) $n=39$, and (c) $n=40$. For all measurements $T_{\mathrm{int}} = 11.5~\mu$s. The continuous red curve in each panel represents the result of a least squares fit performed in the frequency domain (see text for details).}\label{fig3}
\end{figure}

\subsection{Resonances in stronger electric fields}

Additional F\"orster resonances were studied in larger electric fields, and for different values of $n$. To allow transitions to several Rydberg states that ionized in different electric fields to be probed, the depletion of the $|n\mathrm{s}\rangle$ state population was monitored to identify each resonance. The corresponding data, recorded for $T_{\mathrm{int}} = 11.5~\mu$s, are presented in Figure~\ref{fig3}. At $n=40$, transitions from the $|40\mathrm{s}'\rangle$ state to seven $\ell$-mixed Stark states were tuned through resonance with $\nu_{\mathrm{NH}_3}$ in fields between 2.5 and 6~V/cm. A similar set of resonances are seen in Figure~\ref{fig3}(b) for $n=39$. In this case the first strong feature, the $|39\mathrm{s}'\rangle\rightarrow|39\mathrm{p}'\rangle$ transition, occurs in a field of $\sim3.5$~V/cm. In this spectrum, an onset of depletion of the $|39\mathrm{s}\rangle$ signal is also seen in fields below 0.5~V/cm. This is attributed to FRET from the rotational degrees of freedom in NH$_3$ to higher $n$ Rydberg states in He. Although this process affects the background signal level in higher fields, it does not significantly affect the individual resonances of interest here. In Figure~\ref{fig3}(c) a similar onset in the depletion of the $|38\mathrm{s}\rangle$ population occurs in fields up to $\sim0.5$~V/cm. In this case the resonances associated with inversion transitions in NH$_3$ are seen in fields between 5 and 7.5~V/cm. 


The widths of the resonances in Figure~\ref{fig3} were determined by fitting Gaussian functions (continuous red curves) in the frequency domain. For each resonance, the individual field-to-frequency transformation was calculated from the Stark shifts of the Rydberg states. To minimize any impact of the rotational energy transfer, in electric fields $>0.5$~V/cm in Figure~\ref{fig3}(b) and (c), on the analysis, only the parts of each spectrum surrounding the resonances of interest, and encompassed by the fit functions, were considered in the fitting process. 

\begin{figure}
\includegraphics[width=0.85\columnwidth]{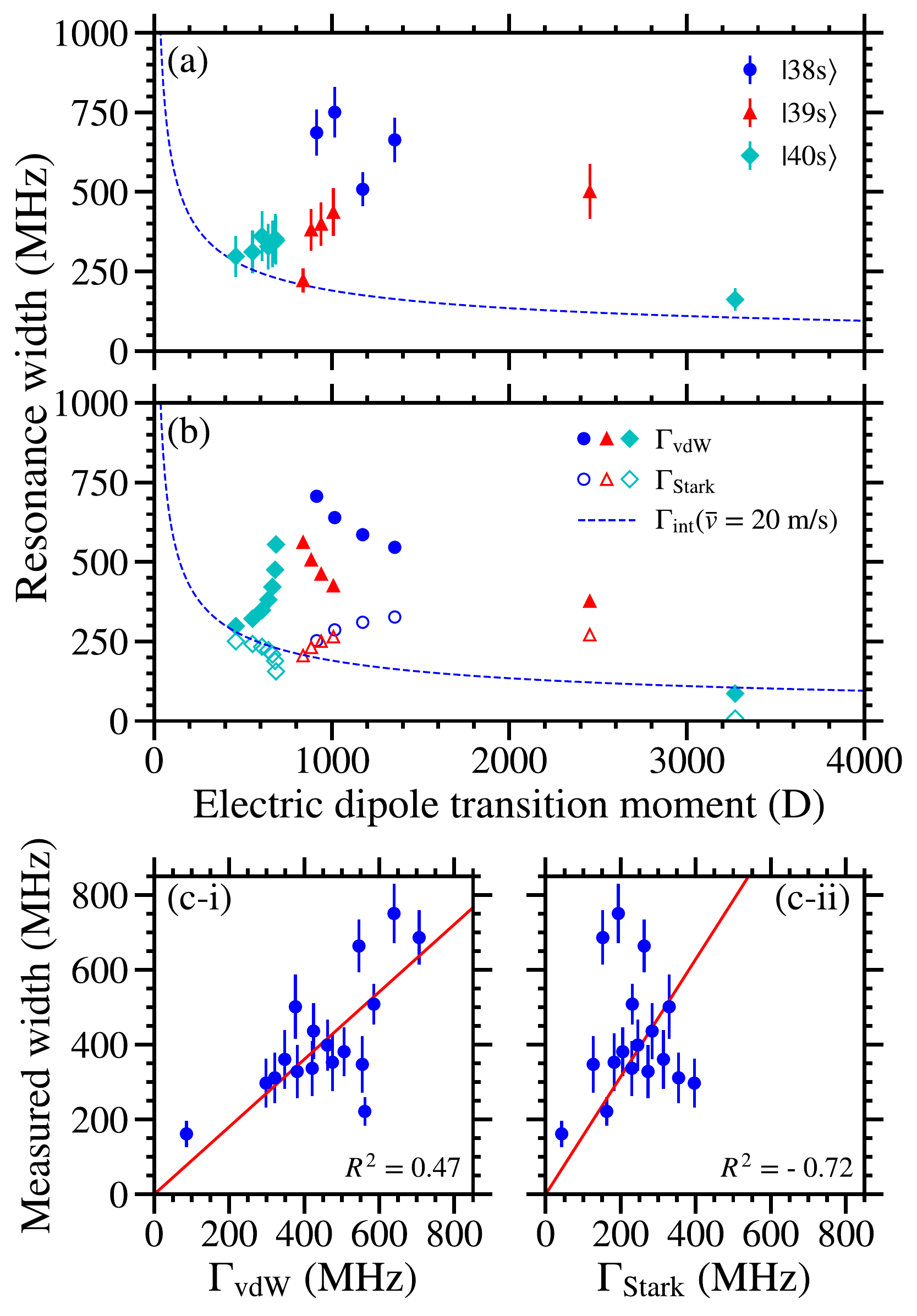}
\caption{(a) Measured FRET resonance widths displayed with respect to the calculated electric dipole moments of the corresponding transitions in He. (b) Resonance widths calculated by considering effects of van~der~Waals interactions between the atoms and molecules undergoing FRET, $\Gamma_{\mathrm{vdW}}$ (filled points), and stray time-varying electric fields, $\Gamma_{\mathrm{Stark}}$ (open points). Calculated widths, $\Gamma_{\mathrm{int}}$, for $\overline{v} = 20$~m/s are displayed as the dashed blue curves in (a) and (b). (c) The correlation between the measured resonance widths and (i) $\Gamma_{\mathrm{vdW}}$, and (ii) $\Gamma_{\mathrm{Stark}}$.}\label{fig4}
\end{figure}

The measured resonance fields in Figure~\ref{fig3} are all in excellent quantitive agreement with those expected from the calculated Stark structure of the Rydberg states. The resonance FWHMs range from 160~MHz to 750~MHz. The relationship between these and the calculated electric dipole moments for the transitions between the Rydberg states, is shown in Figure~\ref{fig4}(a). The resonance widths calculated using Equation~\ref{eq:width} for $\overline{v} = 20$~m/s are indicated in this figure by the dashed blue curve. These widths are similar to those observed in the experiments for the resonances with the smallest ($\sim500$~D) and largest ($\sim3300$~D) values of $\mu_{\mathrm{He}}^{\mathrm{trans}}$. However, between these extrema, there are significant deviations from the measurements. These discrepancies are attributed to effects of static dipole-dipole interactions between the atoms and the molecules as they undergo FRET. 

The induced static electric dipole moments of He atoms in $|n\mathrm{s}'\rangle$ Rydberg states, $\mu_{n\mathrm{s}'}^{\mathrm{static}}$ range from 340~D in the field of 0.78~V/cm associated with the $|40\mathrm{s}'\rangle\rightarrow|40\mathrm{p}'\rangle$ F\"orster resonance, to 2175~D in a field of $\sim5.3$~V/cm. The static dipole-dipole interaction between these and the polar NH$_3$ molecules can broaden the measured resonances. If the angle-average static dipole-dipole interaction energy is considered to represent the standard deviation $s_{\mathrm{vdW}}$ of a Gaussian function, i.e., 
\begin{eqnarray}
s_{\mathrm{vdW}} &=& \frac{\mu_{n\mathrm{s}'}^{\mathrm{static}}\,\mu_{\mathrm{NH}_3}}{4\pi\epsilon_0\,R_{\mathrm{static}}^3},
\end{eqnarray}
where $R_{\mathrm{static}}$ is the typical atom-molecule distance, the resulting FWHM resonance width can be expressed in general as
\begin{eqnarray}
\Gamma_{\mathrm{vdW}} = \frac{\Delta E_{\mathrm{dd}}^{\mathrm{static}}}{hc} &\simeq& s_{\mathrm{vdW}}\,2\sqrt{2\log2}.\label{eq:vdWwidth}
\end{eqnarray}
For the FRET processes considered here, with impact parameters $b$ between 100 and 200~nm, $R_{\mathrm{static}}$ may be considered to be approximately equal to $\langle r_{n\mathrm{s}}\rangle \simeq 1.5\,n^2\,a_0$, the radius of the Rydberg electron charge distribution ($\langle r_{38\mathrm{s}}\rangle = 115$~nm, $\langle r_{39\mathrm{s}}\rangle = 121$~nm, and $\langle r_{40\mathrm{s}}\rangle = 127$~nm). The van~der~Waals widths, $\Gamma_{\mathrm{vdW}}$, calculated for these values of $R_{\mathrm{static}}$ are displayed as the filled points in Figure~\ref{fig4}(b) and follow the general trend exhibited by the experimental data in Figure~\ref{fig4}(a).

Contributions from Stark shifts caused by stray static or time-varying electric fields were identified by recording microwave spectra of Rydberg-Rydberg transitions at a range of times within the interaction electric field pulses. From these spectra (not shown), the maximal temporal variation in field strength within the time $T_{\mathrm{int}}$ was $\pm$1.7\%. These fluctuations originate from imperfect impedance matching, and dominate over spectral broadenings arising from the presence of ions or electrons. Calculated resonance widths $\Gamma_{\mathrm{Stark}}$ that reflect the Stark shift of each transition in these time-varying electric fields, are displayed as the open points in Figure~\ref{fig4}(b). These calculated resonance widths do not follow the trend exhibited by the experimental data, and are all smaller than the experimentally recorded FRET widths. The largest calculated value of $\Gamma_{\mathrm{Stark}}$ is 325~MHz.

To validate the interpretation of the data in Figure~\ref{fig4}(a), the correlation of $\Gamma_{\mathrm{vdW}}$ and $\Gamma_{\mathrm{Stark}}$, with the measured widths are presented in Figure~\ref{fig4}(c-i) and (c-ii), respectively. In the case of $\Gamma_{\mathrm{vdW}}$, the line fit to the data has a slope of $0.90\pm0.06$ indicating a strong direct correlation between the calculated widths and the results of the measurements. The $R^2$ value, describing the correlation of these data with the fit is 0.47, and indicates that the data are reasonably well described by this linear fit function. In contrast, for $\Gamma_{\mathrm{Stark}}$ the slope of the line fit to the data is $1.78\pm0.12$ and $R^2=-0.72$. Therefore, irrespective of the amplitude of the stray time-varying electric fields, there is not a strong correlation between the measured widths and the Stark shifts of the Rydberg-Rydberg transitions.

\section{Conclusion}

The results of the experiments reported here have permitted a detailed study of resonance widths in electric-field-controlled F\"orster resonance energy transfer between Rydberg atoms and polar state state molecules at low translational temperatures. These low temperatures were facilitated, in particular by the reduction of the concentration of the NH$_3$ seed gas in the pulsed supersonic beams used in the experiments, which reduced the velocity slip between the components in the beams. In this setting the FRET process has been exploited to determine $\emph{in situ}$, the mean atom-molecule collision speed, and the molecule number density. 

The correlation of the measured resonance widths in Figure~\ref{fig4}(a) with the spectral broadenings expected from static dipole-dipole interactions between the atoms and the molecules as they undergo FRET leads to the following conclusions: (1) In weak electric fields, because the initial $|n\mathrm{s}'\rangle$ states are not strongly polarized, the FRET resonance widths are dominated by the atom-molecule interaction time, i.e., the center-of-mass collision speed. (2) In stronger fields, when the $|n\mathrm{s}'\rangle$ states are more strongly polarized, the resonances are broadened and the widths are dominated by van~der~Waals interactions. These observations represent essential input for future studies of dipole-bound states of Rydberg atoms and polar ground-state molecules, and for the implementation of proposed schemes for coherent control, cooling and sensing in this hybrid system~\cite{kuznetsova11a,huber12a,zhao12a,zeppenfeld17a}. Investigations of effects of higher order multipole interactions~\cite{patsch22a}, and Rydberg-electron scattering from NH$_3$~\cite{green00a,gonzales21a} will be valuable in achieving a more nuanced interpretation of the data. 

\begin{acknowledgments}
This work was supported by the European Research Council (ERC) under the European Union's Horizon 2020 research and innovation program (Grant No. 683341) and by the UK Science and Technology Facilities Research Council under Grant No. ST/T006439/1.
\end{acknowledgments}


\begin{thebibliography}{99}

\bibitem{zhou21a} H. Zhou, W. E. Perreault, N. Mukherjee, and R. N. Zare, Quantum mechanical double slit for molecular scattering, Science {\bf 374}, 960 (2021).

\bibitem{paliwal21a} P. Paliwal, N. Deb, D. M. Reich, A. van der Avoird, Ch. P. Koch, and E. Narevicius, Determining the nature of quantum resonances by probing elastic and reactive scattering in cold collisions, Nat. Chem. {\bf 13}, 94 (2021).

\bibitem{gao18a} Z. Gao,T. Karman, S. N. Vogels, M. Besemer, A. van der Avoird, G. C. Groenenboom, and S. Y. T. van de Meerakker, Observation of correlated excitations in bimolecular collisions, Nat. Chem. {\bf 10}, 469 (2018).

\bibitem{zou19a} J. Zou, S. D. S. Gordon, and A. Osterwalder, Sub-Kelvin stereodynamics of the Ne($^3$P$_2$) + N$_2$ reaction, Phys. Rev. Lett. {\bf 123}, 133401 (2019).

\bibitem{gawlas20a} K. Gawlas and S. D. Hogan, Rydberg-state-resolved resonant energy transfer in cold electric-field-controlled intrabeam collisions of NH$_3$ with Rydberg He atoms, J. Phys. Chem. Lett. {\bf 11}, 83 (2020).

\bibitem{muller94a} P. Muller, Glossary of terms used in physical organic chemistry (IUPAC Recommendations 1994),  Pure Appl. Chem., {\bf 66}, 1077 (1994).


\bibitem{perrin32a} F. Perrin, Th\'eorie quantique des transferts d'activation entre mol\'ecules de m\^eme esp\`ece  Cas des solutions fluorescentes, Ann Phys. {\bf 10} 283 (1932).

\bibitem{forster46a} Th. F\"orster, Energiewanderung und fluoreszenz, Naturwissenschaften {\bf 33}, 166 (1946).

\bibitem{jones19a} G. A. Jones and D. S. Bradshaw, Resonance energy transfer: From fundamental theory to recent applications, Front, Phys. {\bf 7}, 100 (2019).

\bibitem{gallagher08a} T. F. Gallagher and P. Pillet, Dipole-dipole interactions of Rydberg atoms, Adv. At. Mol. Opt. Phys. {\bf 56} 161 (2008).

\bibitem{zhelyazkova16a} V. Zhelyazkova, R. Jirschik and S. D. Hogan, Mean-field energy-level shifts and dielectric properties of strongly polarized Rydberg gases, Phys. Rev. A {\bf 94}, 053418 (2016).


\bibitem{cederbaum97a} L. S. Cederbaum, J. Zobele, and F. Tarantelli, Giant Intermolecular Decay and Fragmentation of Clusters, Phys. Rev. Lett. {\bf 79}, 4778 (1997).

\bibitem{janke15a} T. Jahnke, Interatomic and intermolecular Coulombic decay: the coming of age story, J. Phys. B: At. Mol. Opt. Phys. {\bf 48} 082001 (2015).

\bibitem{mirkovic17a} T. Mirkovic, E. E. Ostroumov, J. M. Anna, R. van~Grondelle, Govindjee, and G. D. Scholes, Light absorption and energy transfer in the antenna complexes of photosynthetic organisms, Chem. Rev. {\bf 117}, 249 (2017).

\bibitem{blackmore19a} J. A. Blackmore, L. Caldwell, P. D. Gregory, E. M. Bridge, R. Sawant, J. Aldegunde, J. Mur-Petit, D. Jaksch, J. M. Hutson, B. E. Sauer, M. R. Tarbutt and S. L Cornish, Ultracold molecules for quantum simulation: rotational coherences in CaF and RbCs, Quantum Sci. Technol. {\bf 4}, 014010 (2019).


\bibitem{gallagher94a} T. F. Gallagher, \emph{Rydberg Atoms} (Cambridge University Press, Cambridge, U.K., 1994).

\bibitem{farooqi03a} S. M. Farooqi, D. Tong, S. Krishnan, J. Stanojevic, Y. P. Zhang, J. R. Ensher, A. S. Estrin, C. Boisseau, R. C{\^{o}}t{\'{e}}, E. E. Eyler, and P. L. Gould, Long-Range Molecular Resonances in a Cold Rydberg Gas,  Phys. Rev. Lett. {\bf 91}, 183002 (2003).

\bibitem{allmendinger16a} P. Allmendinger, J. Deiglmayr, K. H{\"o}veler, O. Schullian, and F. Merkt, Observation of Enhanced rate coefficients in the H$_{2}^{+}$ + H$_{2} \rightarrow$ H$_{3}^{+}$ + H reaction at low collision energies, J. Chem. Phys. {\bf 145}, 244316 (2016).

\bibitem{kuznetsova11a} E. Kuznetsova, S. T. Rittenhouse, H. R. Sadeghpour, and S. F. Yelin, Rydberg atom mediated polar molecule interactions: A tool for molecular-state conditional quantum gates and individual addressability, Phys. Chem. Chem. Phys. {\bf 13}, 17115 (2011).

\bibitem{huber12a} S. D. Huber and H. P. B\"uchler, Dipole-Interaction-Mediated Laser Cooling of Polar Molecules to Ultracold Temperatures, Phys. Rev. Lett. {\bf 108}, 193006 (2012).

\bibitem{zhao12a} B. Zhao, A. W. Glaetzle, G. Pupillo, and P. Zoller, Atomic Rydberg Reservoirs for Polar Molecules, Phys. Rev. Lett. {\bf 108}, 193007 (2012).

\bibitem{zeppenfeld17a} M. Zeppenfeld, Nondestructive detection of polar molecules via Rydberg atoms, Europhys. Lett. {\bf 118}, 13002 (2017).



\bibitem{smith78a} K. A. Smith, F. G. Kellert, R. D. Rundel, F. B. Dunning and R. F. Stebbings, Discrete energy transfer in collisions of Xe($n$f) Rydberg atoms with NH$_3$ molecules, Phys. Rev. Lett. {\bf 40}, 1362 (1978).

\bibitem{stebbings82a} F. B. Dunning and R. F. Stebbings, Collisions of Rydberg atoms with molecules,  Annu. Rev. Phys. Chem. {\bf 33}, 173 (1982).

\bibitem{petitjean84a} L. Petitjean, F. Gounand and P. R. Fournier, Depopulation of rubidium Rydberg states by CO molecules: An experimental and theoretical study, Phys. Rev. A {\bf 30}, 71 (1984).

\bibitem{petitjean86a} L. Petitjean, F. Gounand and P. R. Fournier, Collisions of rubidium Rydberg-state atoms with ammonia, Phys. Rev. A {\bf 33}, 143 (1986).

\bibitem{ling93a} X. Ling, M. T. Frey, K. A. Smith and F. B. Dunning, The role of inversion transitions in K(150p)/NH$_3$, ND$_3$ collisions, J. Chem. Phys. {\bf 98}, 2486 (1993).

\bibitem{zhelyazkova17a} V. Zhelyazkova and S. D. Hogan, Electrically tuned F\"orster resonances in collisions of NH$_3$ with Rydberg He atoms, Phys. Rev. A  {\bf 95}, 042710 (2017).

\bibitem{zhelyazkova17b} V. Zhelyazkova and S. D. Hogan, Probing resonant energy transfer in collisions of ammonia with Rydberg helium atoms by microwave spectroscopy, J. Chem. Phys. {\bf 147}, 244302 (2017).

\bibitem{jarisch18a} F. Jarisch and M. Zeppenfeld, State resolved investigation of F\"orster resonant energy transfer in collisions between polar molecules and Rydberg atoms, New J. Phys. {\bf 20}, 113044 (2018).



\bibitem{amarasinghe17a} C. Amarasinghe and A. G. Suits, Intrabeam scattering for ultracold collisions, J. Phys. Chem. Lett. {\bf 8}, 5153 (2017).

\bibitem{halfmann00a} T. Halfmann, J. Koensgen, and K. Bergmann, A source for a high-intensity pulsed beam of metastable helium atoms, Meas. Sci. Technol. {\bf 11}, 1510 (2000).

\bibitem{hogan18a} S. D. Hogan, Y. Houston, and B. Wei, Laser photoexcitation of Rydberg states in helium with $n > 400$, J. Phys. B: At. Mol. Opt. Phys. {\bf 51}, 145002 (2018).

\bibitem{drake99a} G. W. F. Drake, High precision theory of atomic helium, Phys. Scr. {\bf T 83}, 83 (1999).

\bibitem{deller18a} A. Deller and S. D. Hogan, Microwave spectroscopy of the 1s$n$p\,$^3$P$_J$ fine structure of high Rydberg states in $^4$He, Phys. Rev. A {\bf 97}, 012505 (2018).

\bibitem{gallagher92a} T.F.Gallagher, Resonant collisional energy transfer between Rydberg atoms, Phys. Rep. {\bf 210}, 319 (1992).

\bibitem{patsch22a} S. Patsch, M. Zeppenfeld and C. P. Koch, Rydberg atom-enabled spectroscopy of polar molecules via Förster resonance energy transfer, arXiv:2205.04327 (2022).

\bibitem{green00a} C. H. Greene, A. S. Dickinson and H. R. Sadeghpour, Creation of Polar and Nonpolar Ultra-Long-Range Rydberg Molecules, Phys. Rev. Lett. {\bf 85}, 2458 (2000).

\bibitem{gonzales21a} R. Gonz\'alez-F\'erez, J. Shertzer and H. R. Sadeghpour, Ultralong-Range Rydberg Bimolecules, Phys. Rev. Lett. {\bf 126}, 043401 (2021).

\end{thebibliography}
\end{document}